%% file: main.tex
\documentclass[fleqn,10pt]{wlscirep}
\usepackage[utf8]{inputenc}
\usepackage[T1]{fontenc}
\usepackage{lineno}
\usepackage{multirow}
\usepackage{graphicx}
\usepackage{subcaption}

\title{Curation and Analysis of MIMICEL – An Event Log for MIMIC-IV Emergency Department}

\author[1,$*$]{Jia Wei}
\author[1]{Chun Ouyang}
\author[1]{Bemali Wickramanayake}
\author[1]{Zhipeng He}
\author[2]{Keshara Perera}
\author[3]{Catarina Moreira}
\affil[1]{School of Information Systems, Queensland University of Technology, Brisbane, 4000, Australia}
\affil[2]{Queensland Health, Brisbane, 4000, Australia}
\affil[3]{Data Science Institute, University of Technology Sydney, Sydney, 2007, Australia}
\affil[$*$]{corresponding author: Jia Wei (jia.wei@hdr.qut.edu.au)}

\begin{abstract}
The global issue of overcrowding in emergency departments (ED) necessitates the analysis of patient flow through ED to enhance efficiency and alleviate overcrowding. However, traditional analytical methods are time-consuming and costly. The healthcare industry is embracing process mining tools to analyse healthcare processes and patient flows. Process mining aims to discover, monitor, and enhance processes by obtaining knowledge from event log data. However, the availability of event logs is a prerequisite for applying process mining techniques. Hence, this paper aims to generate an event log for analysing processes in ED. In this study, we extract an event log from the MIMIC-IV-ED dataset 
and name it MIMICEL. MIMICEL captures the process of patient journey in ED, allowing for analysis of patient flows and improving ED efficiency. We present analyses conducted using MIMICEL to demonstrate the utility of the dataset. The curation of MIMICEL facilitates extensive use of MIMIC-IV-ED data for ED analysis using process mining techniques, while also providing the process mining research communities with a valuable dataset for study.

\end{abstract}
\begin{document}

\flushbottom
\maketitle

\thispagestyle{empty}

\input{background_summary}

\input{methods} 
\input{data_records}

\input{technical_validation} 
\input{usage_notes}

\input{code_availability}

\input{Appendix}

\end{document}

%% file: background_summary.tex

\section*{Background \& Summary}

Emergency departments (ED) are one of the most important hospital departments~\cite{ibanez2021interactive}. Due to the variety of diseases and the urgency of the patients, ED processes are complex and involve various activities requiring multidisciplinary human and medical resources~\cite{Duma2018AnAH}. This has also contributed to ED overcrowding, a widespread problem in the global healthcare sector. The issue is largely caused by the inability of emergency services to meet rising demand~\cite{Savioli2022EmergencyDO}, which can further compromise the quality and accessibility of healthcare services~\cite{Duma2018AnAH}. Therefore, it is necessary to analyse the flow of patients through ED to increase the effectiveness of the processes and reduce overcrowding~\cite{brenner2010modeling}. In general, healthcare processes, such as the process of patient activities during their stay in an emergency department (referred to as the ED process), are regarded as dynamic, complex, multidisciplinary and ad-hoc~\cite{rebuge2012business}. In the absence of a comprehensive view of the end-to-end process, traditional analysis methods, such as interviews and group meetings to obtain insight into the process, is time-consuming, costly, and subjective~\cite{delias2014applying} when applied to ED process analysis.

An increasing number of research works have investigated process mining for the healthcare domain~\cite{delias2014applying}, with an aim to analyse healthcare process performance using process execution data recorded in the health information systems. Munoz-Gama et al.~\cite{MunozGama2022ProcessMF} outline the distinct characteristics of healthcare processes and associated challenges that need to be addressed when utilising process mining for healthcare process analysis. In addition, Martin et al.~\cite{Martin2020RecommendationsFE} provide recommendations for process mining researchers and stakeholders in the healthcare domain, respectively, for applying process mining to healthcare to enhance the usability and comprehension of these applications. In particular, process mining techniques are increasingly adopted to analyse patient flows in the ED. Delias et al.~\cite{delias2014applying} demonstrate the use of process discovery techniques to identify and analyse ED processes. The discovered process model can visualise the patient's path through the emergency department. As a result of the visualisation of ED processes, process knowledge, such as activity frequency and process patterns, can be captured and used for process performance analysis (e.g., to identify bottlenecks that affect process efficiency). Similarly, Cho et al.~\cite{cho2020process} introduce a process performance indicator framework for managing emergency room processes, assessing performance based on four dimensions: time, cost, quality, and flexibility.

Event log data is the foundation of process mining and is usually represented in the form of sequential tabular data. An event log is a collection of \textit{cases}, and each case consists of a sequence of \textit{events} (ordered according to when they occurred)~\cite{vanderAalst2016ProcessMD}. The availability of the event log is a prerequisite for applying process mining techniques. This work aims to extract event logs from the MIMIC-IV-ED~\cite{mimiced} dataset, an extensive, freely available database containing unidentifiable health-related data from Beth Israel Deaconess Medical Center. The MIMIC-IV-ED dataset contains data tables that capture individual patient activities during the ED process and are linked using an existing relational database schema. Although these data tables provide a snapshot of the patient journey, they do not depict the patient's end-to-end process in the ED. 
To comprehend and analyse the ED process, we follow a well-established guideline~\cite{guideline} for generating event logs 
and name this log \textbf{MIMICEL}. The extracted \textbf{MIMICEL} is intended to capture an end-to-end patient journey, facilitating the analysis of existing patient flows to enhance the efficiency of the ED process. Furthermore, the curation of \textbf{MIMICEL} makes the MIMIC-IV-ED data accessible in the form of an event log, which enables the use of process mining techniques to analyse ED processes. It also provides the process mining research communities with a valuable dataset for study.

%% file: methods.tex
\section*{Methods}

In existing research, various methods have been proposed to generate event logs. Remy~et~al.~\cite{remy2020event}
introduce a method that uses structured data from data warehouses and clinical guidelines to identify process-related data. This approach requires significant manual effort, including domain expert consultations, and is heavily based on domain-specific knowledge. 
Rojas~et~al.~\cite{rojas2017question} propose a method to extract data from hospital information systems to generate event logs for analysing ED processes. However, their approach relies on predefined expert queries to determine the data to be included in the event logs. Andrews~et~al.~\cite{andrews2020quality} introduce a semi-automatic, domain-independent method for event log generation, 
focusing on integrating data quality assessment metrics in log generation. 
However, both methods lack a systematic procedure for creating event logs, which limits their reproducibility.

\begin{figure}[b!!!]
    \centering
    \includegraphics[width=.9\textwidth]{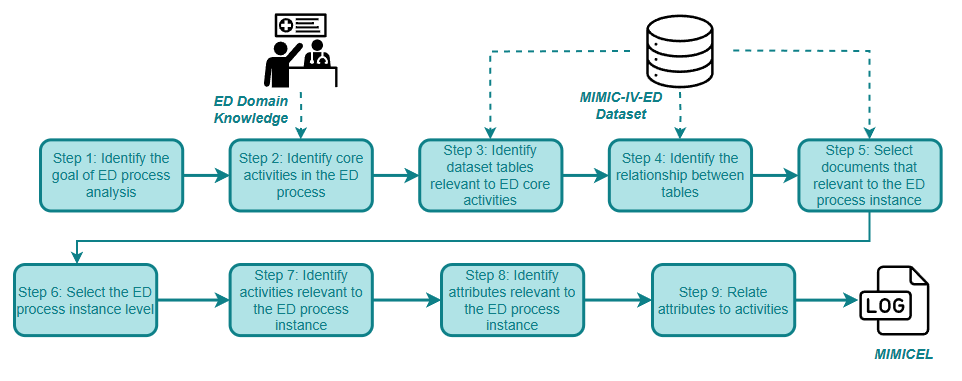}
    \caption{An overview of the method for generating MIMICEL}
    \label{fig:Figure1}
\end{figure}

In this study, we adhere to the guideline proposed by Jans et al.~\cite{guideline}, which is considered the most comprehensive and systematic approach for the extraction of event logs. 
It consists of nine steps that we follow to extract event logs specifically tailored to the objective of analysing ED processes. 
Below, we discuss the method used to extract an event log that captures the execution of ED processes from the MIMIC-IV-ED dataset~\cite{mimiced}. Figure~\ref{fig:Figure1} depicts an overview of the method.


\begin{itemize}
    \item \textbf{\textit{Step 1: Identify the goal of ED process analysis}}
    
    This step focuses on defining the objective of event log generation, which needs to be aligned with the requirements of the project sponsor~\cite{guideline}. ED overcrowding remains a critical issue in healthcare, driven by the growing demand for emergency care and limited service availability~\cite{Savioli2022EmergencyDO}. Its consequences include
    prolonged patient waiting times, patients leaving without being treated, poor quality of care provided to patients and the high stress placed on emergency department staff~\cite{Duma2018AnAH}. Addressing these challenges by improving ED patient flow is a priority for healthcare providers~\cite{Andrews2018ImprovingPF}. 
    
    To improve ED process efficiency, it is essential to first understand a patient's journey within ED. This has motivated the main objective of this work, which is to capture an end-to-end patient journey using data from the MIMIC-IV-ED dataset.

    \item \textbf{\textit{Step 2: Identify core activities in the ED process}} 
    
    This step involves defining the process boundaries and identifying core activities relevant to stakeholders~\cite{guideline}, guided by domain knowledge. 
     In this work, the widely-adopted conceptual model of emergency department crowding~\cite{Asplin2003ACM} is referenced as the basis for identifying key activities in the ED process. This model identifies three interdependent components—input, throughput, and output—as contributors to ED crowding, emphasising the necessity of examining the entire acute care system to address the issue effectively. However, as this work focuses on activities within the ED, only the throughput component is considered.
     
    The throughput component highlights the internal process in the ED, which is comprised of two primary phases. Upon arrival at the ED, patients would be triaged and placed in different rooms, constituting the first phase. Next, the patient would receive some diagnostic tests and treatment, mainly in the second phase. To this end, some patients may leave the ED without being treated completely. 
    Based on consultations with an ED doctor, it was confirmed that the ED operates independently of other hospital services. Therefore, this work focuses on the activities described in the throughput component of the model~\cite{Asplin2003ACM} as key process cornerstones specific to the ED:
    \begin{itemize}
        \setlength\itemsep{0em}
        \item \textit{Patient arrives at ED}
        \item \textit{Triage and room placement}
        \item \textit{Diagnostic evaluation and ED treatment}
        \item \textit{Patient disposition}
    \end{itemize}

    \item \textbf{\textit{Step 3: Identify tables in the datasets that reflect ED core activities}} 
    
    In this step, the key process activities identified in Step~2 serve as a guide for selecting key tables. In our work, all tables within the MIMIC-IV-ED dataset are essential for representing the core activities of an ED process, and therefore, all tables are included. Table~\ref{table1} provides a detailed mapping between the cornerstone activities and their corresponding tables in the MIMIC-IV-ED dataset. 
    
    \begin{table}[h]
    \begin{center}
    \resizebox{0.6\textwidth}{!}{%
    \begin{tabular}{@{}cc@{}}
    \hline
    Cornerstones & Tables \\
    \hline
    Patient arrives at ED   & \textit{edstays} \\
    Triage and room placement    & \textit{triage} \\
    Diagnostic evaluation and ED treatment   & \textit{diagnosis} \& \textit{medrecon} \& \textit{vitalsigns} \& \textit{pyxis}  \\
    Patient disposition   & \textit{edstays} \\
    \hline
    \end{tabular}
    }
    \caption{Mappings between cornerstones and key tables in the MIMIC-IV-ED dataset}
    \label{table1}
    \end{center}
    \end{table}
    
    \item \textbf{\textit{Step 4: Identify relationships between tables}} 
    
    This step focuses on identifying  the relationships among the tables selected in Step~3. In this work, the relationships between these tables are defined based on existing relational database schema of the MIMIC-IV-ED dataset. To illustrate these relationships, an Entity-Relationship (ER) diagram is provided in Figure~\ref{fig:Figure2}.
    
    \begin{figure*}[h]
        \centering
        \includegraphics[width =.95\textwidth]{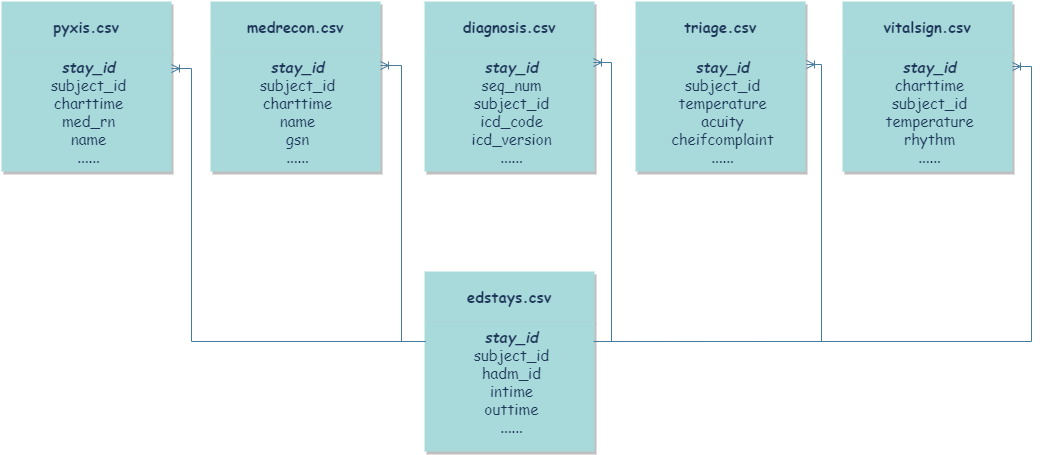}
        \caption{Relationships between key tables in the MIMIC-IV-ED dataset}
        \label{fig:Figure2}
    \end{figure*}
    
    \item \textbf{\textit{Step 5: Select documents relevant to ED process instance}}
    
   This step aims to define the boundaries of process instances (i.e., cases) by identifying the start document that triggers an instance of the process or the end document that signifies the completion of the process instance. In this work, the temporal information available in most tables of the MIMIC-IV-ED dataset supports the identification of start and end activities within the ED process.
    
    For the start of an ED stay, various activities may occur. If the patient's arrival at ED is the initial event, the \textbf{edstays} table which ``tracks patient admissions to the ED'' 
    is considered the start document that triggers the process. Alternatively, if routine vital signs measurement, medicine reconciliation or medicine dispensation occurs prior to the patient's arrival at ED (e.g., in an ambulance), the corresponding table \textbf{vitalsign}, \textbf{medrecon} or \textbf{pyxis} are used as the start documents. 
    
    At the end of an ED stay, patients are discharged from the ED. Information regarding the patient's diagnoses, which is used for billing purposes, is recorded in the \textbf{diagnosis} table. This table serves as the end document, marking the completion of a patient's ED process.
    
    \item \textbf{\textit{Step 6: Select the ED process instance level}}
    
    This step determines the granularity of case in the extracted event log. In line with the primary objective of this study, the focus is on individual ED stays, each uniquely identified by \textit{stay\_id}, which serves as the \textit{case ID} in the extracted event log. The process instance documents identified in Step~5 represent the start and end activities of a single ED stay.  It is important to note that a patient may have multiple ED stays, each treated as a separate process instance.

    \item \textbf{\textit{Step 7: Identify activities relevant to ED process instances}}

    This step focuses on selecting relevant activities for the case level (identified in Step~6) based on the data recorded in the key tables identified in Step~3. In this study, all potential activities within a single ED stay are identified using data from the MIMIC-IV-ED dataset.
    
    According to the guideline~\cite{guideline}, candidate activities are expected to have temporal information stored in the database. Although \textit{triage} is a key activity in ED processes~\cite{Asplin2003ACM}, the triage table in the MIMIC-IV-ED dataset does not provide timestamps. Based on the dataset documentation, ``the closest approximation to triage time is the \textit{intime} of the patient from the \textbf{edstays} table''~\cite{mimiced}. 
    Hence, we assign an artificial timestamp to the triage activity by adding ``one second'' to the time when the patient enters the ED (i.e., \textit{intime} of the \textbf{edstays} table). This adjustment ensures that the triage activity does not affect the time of any subsequent activities. 
   
    Table~\ref{table2} lists the identified activities relevant to ED process instances and their corresponding temporal information from the relevant tables in the MIMIC-IV-ED dataset. At the end of this step, we have identified three mandatory attributes of an event log, i.e., \textit{case ID}, \textit{activity name} and \textit{timestamp}. Since the discharge time of an ED visit cannot precede or coincide with its entry time, cases violating this rule were filtered out.
    
    \begin{table}[!ht]
    \begin{center}
    \resizebox{0.6\textwidth}{!}{%
    \begin{tabular}{ccc}
    \hline
    Activity name  & Tables & Column for time information \\
    \hline
    Enter the ED & \textit{edstays} & intime \\
    Triage in the ED  & \textit{edstays} & (intime + 1 second) \\
    Vital sign check  & \textit{vitalsign} & charttime \\
    Medicine reconciliation  & \textit{medrecon} & charttime \\
    Medicine dispensation  & \textit{pyxis} & charttime \\
    Discharge from the ED & \textit{edstays} & outtime \\
    \hline
    \end{tabular}
    }
    \caption{Activities and their time information stored in the MIMIC-IV-ED dataset}
    \label{table2}
    \end{center}
    \end{table}
    
    \item \textbf{\textit{Step 8: Identify attributes relevant to ED process instance}}
    
    In this step, the objective is to identify all relevant attributes in addition to the three mandatory attributes of an event log, based on the available data. In this work, all data attributes stored in the MIMIC-IV-ED dataset are included as relevant attributes.
    
    \item \textbf{\textit{Step 9: Relate attributes to activities}}

    This step aims to relate the attributes identified in Step~8 to either a case or to an event within an event log, i.e., case attribute or event attribute. An ED event log is then extracted based on these mappings. The following tables (Table~\ref{tab:table3} \& Table~\ref{tab:table4}) present a comprehensive description of the attributes in the event log and their corresponding categories. Timestamps listed in Table \ref{table2} are excluded from these tables. At this stage, \textbf{MIMICEL} has been extracted.

    \begin{table}[!h]
    \centering
    \resizebox{.8\textwidth}{!}{%
    \begin{tabular}{cccc}
    \hline
    \multicolumn{1}{l}{Type} & Name & Descriptions & Tables   \\ \hline
    \multirow{8}{1.5cm}{Case Attributes} & stay\_id  & an identity that specifies each emergency department stay & \textit{edstays}     \\
    & subject\_id    & an identity that specifies an individual patient & all tables \\
    & gender & a patient's administrative gender & \textit{edstays}\\
    & race & a patient's self-reported race & \textit{edstays} \\
    & arrival\_transport & a method describes how a patient arrived at the ED & \textit{edstays}\\
    & disposition & a destination after patient getting discharged from the ED & \textit{edstays} \\
    & acuity & a patient's acuity level that was assigned by triage nurse  & \textit{triage} \\
    & chiefcomplaint & a de-identified free-text description of the patient's  chief complaint & \textit{triage} \\ \hline
    \end{tabular}%
     }
    \caption{Descriptions of case attributes in the event log extracted from the MIMIC-IV-ED dataset}
    \label{tab:table3}
    \end{table}
    \begin{table}[h!!!!]
    \resizebox{\textwidth}{!}{%
    \begin{tabular}{cccc}
    \hline
    \multicolumn{1}{l}{Type} & Name & Descriptions & Tables        \\ \hline
    \multirow{22}{1.5cm}{Event Attributes} & hadm\_id       & an identity that specifies a single hospital admission & \textit{edstays}          \\
    & temperature    & a patient's temperature  & \textit{triage}, \textit{vitalsign} \\
    & heartrate      & a patient's heart rate & \textit{triage}, \textit{vitalsign} \\
    & resprate       & a patient's respiratory rate & \textit{triage}, \textit{vitalsign} \\
    & o2sat          & a patient's peripheral oxygen saturation rate & \textit{triage}, \textit{vitalsign} \\
    & sbp            & a patient's systolic blood pressure & \textit{triage}, \textit{vitalsign} \\
    & dbp            & a patient's diastolic blood pressure & \textit{triage}, \textit{vitalsign} \\
    & pain           & a patient's self-reported pain level & \textit{triage}, \textit{vitalsign} \\
    & rhythm         & a patient's heart rhythm & \textit{vitalsign}      \\
    & med\_rn        & a number which indicates how many times of dispensations for a patient & \textit{pyxis}         \\
    & seq\_num       & a number which indicates how many diagnoses were provided to the patient & \textit{diagnosis} \\
    & name  & name of medications & \textit{medrecon}, \textit{pyxis}   \\
    & gsn  & the Generic Sequence Number for the medication & \textit{medrecon}, \textit{pyxis}   \\
    & ndc            & the National Drug Code for the medication & \textit{medrecon}          \\
    & etc\_rn        & a sequential number for differentiating medications in different Enhanced Therapeutic Class (ETC) groups & \textit{medrecon}      \\
    & etccode        & a code for the belonging ETC group & \textit{medrecon}        \\
    & etcdescription & descriptions about the ETC group & \textit{medrecon}          \\
    & gsn\_rn        & a sequential number for differentiating medications in different Generic Sequence Number(GSN) groups & \textit{pyxis}            \\
    & icd\_code      & an International Classification of Diseases (ICD) code that represents the diagnosis & \textit{diagnosis}       \\
    & icd\_version   & the version of ICD system related & \textit{diagnosis}        \\
    & icd\_title     & the textual description of the diagnosis \textit{diagnosis}         \\ \hline
    \end{tabular}%
     }
    \caption{Descriptions of event attributes in the event log extracted from the MIMIC-IV-ED dataset}
    \label{tab:table4}
    \end{table}

\end{itemize}



%% file: data_records.tex
\section*{Data Records}



In this work, the extracted event log \textbf{MIMICEL} is provided in two different formats: one as a CSV file and another in XES (Extensible Event Stream) format~\cite{xes_standard}. Table~\ref{tab:event_log_characteristics} provides a summary of \textbf{MIMICEL} data statistics. The \textbf{MIMICEL} dataset has been published on 
PhysioNet~\cite{physionet}. Details about the dataset can be found via~\href{https://physionet.org/content/mimicel-ed/2.1.0/}{https://physionet.org/content/mimicel-ed/2.1.0/}. 

\begin{table}[!h]
\centering
\resizebox{.4\textwidth}{!}{%
\begin{tabular}{cc}
\hline
\textbf{Attribute}       & \textbf{Value} \\ \hline
\# Cases ({\it stay\_id})  &    425,028     \\
\# Patients ({\it subject\_id}) & 205,466 
\\
\# Events                &   7,568,824    \\
\# Activity types          &        6       \\
Avg. \# events per case &        18      \\
Min. \# events per case ({\it the shortest case})       &        3       \\
Max. \# events per case ({\it the longest case})        &       218      \\
\hline
\end{tabular}%
}
\caption{Descriptive statistics of MIMICEL}
\label{tab:event_log_characteristics}
\end{table}

The extracted \textbf{mimicel.csv} contains in total 7,568,824 events and 425,028 cases recording the ED stays of 205,466 patients in the MIMIC-IV-ED dataset~\cite{mimiced}. Each row in the CSV file represents an execution of an event during an ED stay, and each column corresponds to an attribute of that event. The \textbf{MIMICEL} dataset has three mandatory attributes represented by three columns, namely \textit{stay\_id} (i.e., case ID of \textbf{MIMICEL}), \textit{activity} and \textit{timestamps}. Other columns in \textbf{MIMICEL} represent case attributes and event attributes. Descriptions of all columns representing case attributes and event attributes are provided in Table~\ref{tab:table3} and Table~\ref{tab:table4}.

Table~\ref{tab:table_eventlog} provides a snippet of \textbf{MIMICEL}, illustrating three cases identified by their respective \textit{stay\_id} 35146496, 32354539 and 30505340. These cases are associated with the same \textit{subject\_id} (i.e., 10010848), demonstrating that a single patient may have multiple ED visits. Events within each case are ordered according to their timestamps. Each ED visit is characterised by distinct case attributes; in this snippet, these attributes include \textit{arrival\_transport}, \textit{disposition} and \textit{acuity}. For example, the ED visit with \textit{stay\_id} 35146496 arrived at the ED through the ambulance and was sent home after discharge from the ED. These cases also have event attributes such as \textit{temperature}, \textit{pain} and \textit{seq\_num}, of which the values may change along with the execution of events. 

\begin{table}[!h]
\resizebox{\textwidth}{!}{%
\begin{tabular}{|c|c|c|c|c|c|c|c|c|c|}
\hline
\textbf{stay\_id} & \textbf{subject\_id} & \textbf{timestamp}  & \textbf{activity}       & \textbf{arrival\_transport} & \textbf{disposition} & \textbf{temperature} & \textbf{pain} & \textbf{acuity} & \textbf{seq\_num} \\ \hline
35146496          & 10010848             & 31.10.2165 11:33:00 & Enter the ED            & AMBULANCE                   &                      &                      &               &                 &                   \\ \hline
35146496          & 10010848             & 31.10.2165 11:33:01 & Triage in the ED        &                             &                      & 98.8                 & 0             & 2               &                   \\ \hline
35146496          & 10010848             & 31.10.2165 11:36:00 & Vital sign check        &                             &                      & 98.8                 & 0             &                 &                   \\ \hline
35146496          & 10010848             & 31.10.2165 11:51:00 & Medicine reconciliation &                             &                      &                      &               &                 &                   \\ \hline
35146496          & 10010848             & 31.10.2165 13:45:00 & Vital sign check        &                             &                      &                      &               &                 &                   \\ \hline
35146496          & 10010848             & 31.10.2165 13:58:00 & Discharge from the ED   &                             & HOME                 &                      &               &                 & 1                 \\ \hline
32354539          & 10010848             & 26.06.2169 18:16:00 & Enter the ED            & WALK IN                     &                      &                      &               &                 &                   \\ \hline
32354539          & 10010848             & 26.06.2169 18:16:01 & Triage in the ED        &                             &                      & 97.9                 & 2             & 2               &                   \\ \hline
32354539          & 10010848             & 26.06.2169 22:28:00 & Medicine reconciliation &                             &                      &                      &               &                 &                   \\ \hline
32354539          & 10010848             & 26.06.2169 23:03:00 & Medicine dispensation   &                             &                      &                      &               &                 &                   \\ \hline
32354539          & 10010848             & 27.06.2169 01:43:00 & Discharge from the ED   &                             & HOME                 &                      &               &                 & 1                 \\ \hline
30505340          & 10010848             & 26.01.2170 18:26:00 & Enter the ED            & WALK IN                     &                      &                      &               &                 &                   \\ \hline
30505340          & 10010848             & 26.01.2170 18:26:01 & Triage in the ED        &                             &                      & 98.2                 & 6             & 2               &                   \\ \hline
30505340          & 10010848             & 26.01.2170 18:28:00 & Vital sign check        &                             &                      & 98.2                 & 6             &                 &                   \\ \hline
30505340          & 10010848             & 26.01.2170 22:48:00 & Medicine reconciliation &                             &                      &                      &               &                 &                   \\ \hline
30505340          & 10010848             & 26.01.2170 22:53:00 & Vital sign check        &                             &                      & 98.0                 & 6             &                 &                   \\ \hline
30505340          & 10010848             & 26.01.2170 23:47:00 & Medicine dispensation   &                             &                      &                      &               &                 &                   \\ \hline
30505340          & 10010848             & 27.01.2170 00:42:00 & Vital sign check        &                             &                      & 98.1                 & 4             &                 &                   \\ \hline
30505340          & 10010848             & 27.01.2170 00:44:00 & Discharge from the ED   &                             & HOME                 &                      &               &                 & 1                 \\ \hline
\end{tabular}%
}
\caption{A snippet of extracted \textbf{MIMICEL} in CSV format}
\label{tab:table_eventlog}
\end{table}
\begin{figure}[h!]
    \centering
    \includegraphics[width=0.63\textwidth]{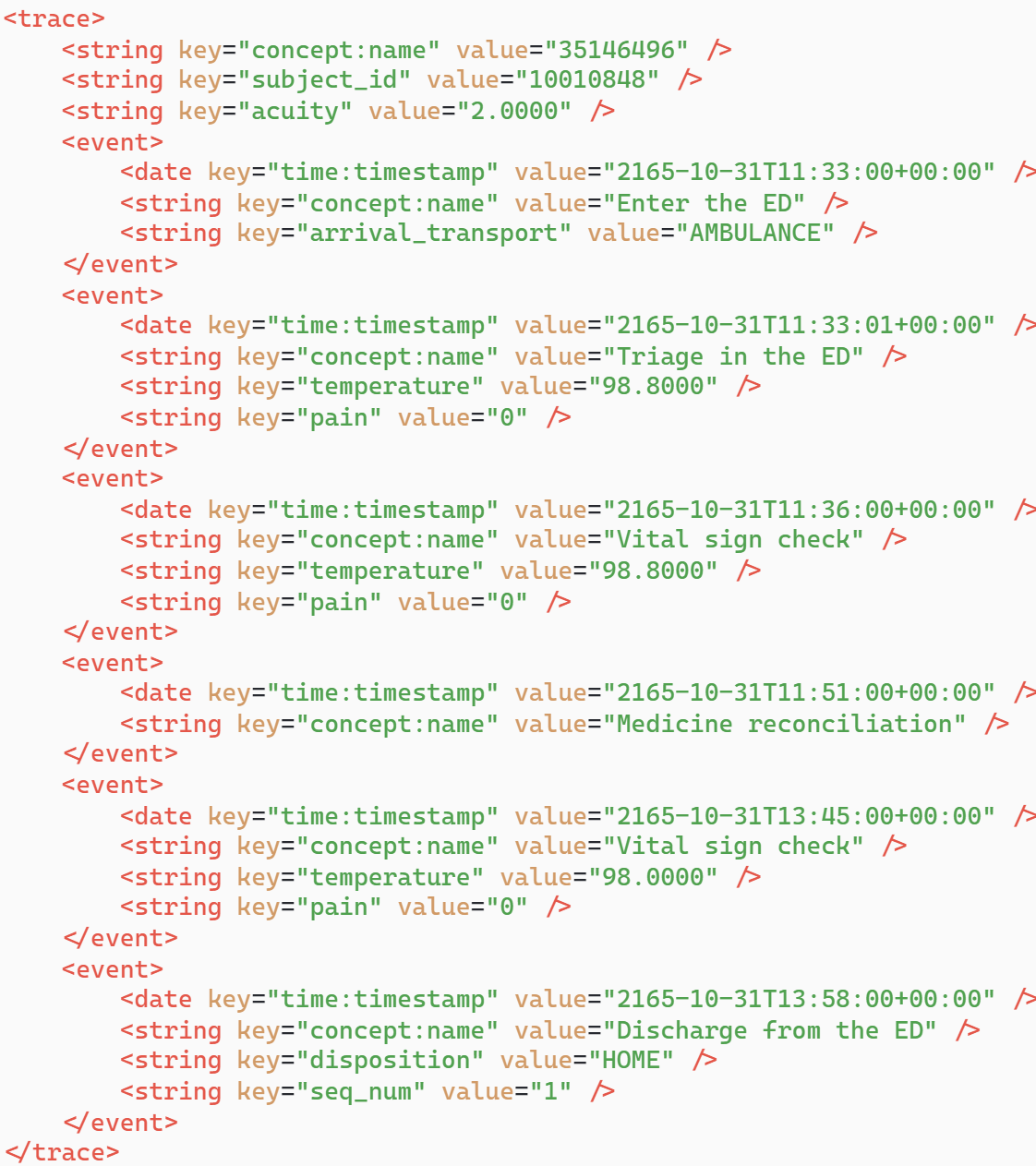}
    \caption{A snippet of extracted MIMICEL in XES format}
    \label{fig:mimicel-xes}
\end{figure}

The \textbf{mimicel.xes} file applies a standard XML-based format for event logs, known as XES (which stands for ``eXtensible Event Stream'')~\cite{xes_standard}. XES maintains the general structure of an event log and uses the term ``trace'' instead of ``case''. This format is widely supported by process mining tools. The \textbf{mimicel.xes} file is created by converting from the original \textbf{mimicel.csv} file using the Python library PM4PY~\cite{Berti2019ProcessMF}. In the XES file, the case attributes from the CSV file are transformed into the trace attributes. An example snippet of the XES event log is shown in Figure~\ref{fig:mimicel-xes}, which corresponds to the example provided in Table~\ref{tab:table_eventlog}.

%% file: technical_validation.tex
\section*{Technical Validation}



In this section, we validate the quality of the extracted \textbf{MIMICEL} to ensure that no errors are introduced when the log extraction is performed and that the extracted log is reliable for analysing ED processes. 
The validation follows the quality assessment framework proposed by Vanbrabant et al.~\cite{vanbrabant2019quality}, which integrates the taxonomy of Bose et al.~\cite{Bose2013WannaIP} to categorise quality issues in event logs and specifically addresses data quality issues in ED medical records. 
Vanbrabant et al.~\cite{vanbrabant2019quality} also introduce DaQAPO\cite{martin2022daqapo}, an R-based tool designed for systematic quality assessment as an implementation of their framework. 
We use DaQAPO to assess the quality of \textbf{MIMICEL}, with details of the coding implementation provided in the Code Availability Section. 
The results, summarised in Table~\ref{tab:tvresults}, are discussed below. 
\begin{table}[]
\resizebox{\textwidth}{!}{%
\begin{tabular}{|c|c|c|}
\hline
\textbf{Data quality problems}                    & \textbf{Items for validation}                                                                                  & \textbf{Results of validation}                              \\ \hline
Missing values                                    & attribute \textit{acuity}                                                                     & 1.64\%                                                      \\ \hline
\multirow{2}{*}{Violation of mutual dependencies} & attribute \textit{disposition} = ``ADMITTED'' \& \textit{hadm\_id} is ``NA'' & 0.24\%                                                      \\ \cline{2-3} 
                                                  & attribute \textit{disposition} = ``HOME'' \& \textit{hadm\_id} is not ``NA'' &  15.1\%      \\ \hline
Outside domain range                              & attribute \textit{pain}                                                                       & 29.4\%                                                      \\ \hline
\multirow{2}{*}{Inconsistent formatting}          & attribute \textit{timestamp}                                                                  & accurate to different time granularities (minute or second) \\ \cline{2-3} 
                                                  & attribute \textit{temperature}                                                                & represented in Fahrenheit or Celsius                        \\ \hline
\end{tabular}%
}
\caption{Data quality issues identified in MIMICEL}
\label{tab:tvresults}
\end{table}

\paragraph{Missing values} This refers to data values that should have been recorded in the event log but are absent. In \textbf{MIMICEL}, certain attributes, including \textit{subject\_id}, \textit{gender} and \textit{race}, \textit{acuity}, \textit{arrival\_transport}, and
\textit{disposition} are mandatory for each case. Assessing these attributes for missing values is crucial for ensuring data completeness. To identify missing values of these attributes, we employ the  function~\textit{missing\_values} from DaQAPO, which detects missing values at different levels of granularity, including activity and specified attribute columns~\cite{martin2022daqapo}. As a result, we detected 1.64\% of cases lacking a value for the attribute \textit{acuity}.

It is essential to distinguish between missing values and null values, as null values can be valid in specific contexts. For example, the attribute \textit{hadm\_id} (hospital admission ID) will be null if a patient is not admitted to the hospital (e.g. when the \textit{disposition} is ``ADMITTED''). 
In addition, the number of incomplete patient ED visit records serves as an essential quality metric~\cite{vanbrabant2019quality}. 
Each ED visit record in the MIMIC-IV-ED dataset is expected to include mandatory activities such as entering, triage, and discharge. Missing one or more of these activities indicates an incomplete ED visit. To detect such cases, we utilise the \textit{incomplete\_cases} function, which identifies records with missing one or more mandatory activities. 


\paragraph{Violation of mutual dependencies} 
This issue refers to the violations in the mutual dependencies between elements within the event log, such as between activities, attributes, or between activities and attributes. For example, the MIMIC-IV-ED dataset documentation specifies a dependency between \textit{hadm\_id} and \textit{disposition}. If a patient is admitted to the hospital after discharge (i.e., \textit{disposition} = "ADMITTED"), the \textit{hadm\_id} should include the corresponding hospital identifier. Conversely, if a patient is discharged home (i.e., \textit{disposition} = "HOME"), the \textit{hadm\_id} should not contain a hospital identifier.

Violations of these dependencies are identified using the \textit{attribute\_dependencies} function. The results reveal that 99.76\% of cases where patients were admitted to the hospital have a valid \textit{hadm\_id}. However, 15.1\% of cases in which patients were discharged home incorrectly retain a record of \textit{hadm\_id}.

\paragraph{Invalid timestamp} This quality issue relates to the validity of the recorded timestamps. To detect invalid timestamps, the \textit{time\_anomalies} function was applied to detect zero or negative case durations. The validation result indicates that no cases in \textbf{MIMICEL} exhibit invalid case duration. This observation also suggests that there is no violation of the logical order within the event log, as the "Enter" activity always occurs before the "Discharge" activity.


Repeated activities with identical timestamps may appear to indicate imprecise timestamps~\cite{vanbrabant2019quality}. To identify such cases, we apply the \textit{multi\_registration} function, which detects activities repeated at the same timestamp within a case. The validation results are summarised as follows:
\begin{itemize}
    \item Out of total 425,028 cases in \textbf{MIMICEL}, 304,369 cases have the activity ``Medicine reconciliation'' executed. Within this subset of cases, 87.07\% (265,016 out of 304,369) exhibit repeated instances of the ``Medicine reconciliation'' activity occurring at the same point in time. This occurs due to unique medication information being recorded for each reconciliation, even when performed simultaneously.
    \item ``Medicine dispensation'' was performed in 295,998 out of 425,028 cases in the \textbf{MIMICEL}. In this subset of cases, 68.84\% (203,770 out of the 295,998) have repeated instances of the activity ``Medicine dispensation'' occurring at the same point in time. These repetitions arise because each occurrence represents the dispensation of a distinct medication, even when multiple medications are dispensed at once.
    \item 60.13\% (256,308 of the 425,028) of cases contain multiple instances of the activity ``Discharge from the ED'' executed at the same time. This happens because each occurrence is tied to a distinct diagnosis, and a single discharge event may involve multiple diagnoses.
\end{itemize}
In \textbf{MIMICEL}, these repeated activities with identical timestamps are valid and not considered a quality issue. These repetitions occur because their corresponding attributes capture distinct details specific to each instance. The event log format (i.e., XES standard) allows only one value per attribute, requiring these repetitions to accurately capture the granularity of event-specific information. As a result, these repetitions highlight the richness of the data rather than indicating any issue in data quality.

\paragraph{Outside domain range} This refers to the identification of attribute values that fall outside the range of possible values. According to the description of the MIMIC-IV-ED dataset, 
patients' self-reported pain levels are expected to be within a range of 0-10. Therefore, when detecting violations of the value range for these specific attributes, any values outside the range of 0-10 for pain level would be considered violations. To identify attributes with value that is out of domain range, we use the function \textit{attribute\_range}. The results reveal that 29\% of cases contain values for \textit{pain} that fall outside the range of 0 to 10.


\paragraph{Inconsistent formatting} This issue concerns the data values that do not conform to a consistent format. 
Consequently, timestamps of activities in \textbf{MIMICEL} such as ``Medicine reconciliation'', ``Medicine dispensation'' and ``Vital sign check'' are accurate to minute level. The timestamps of activities ``Enter the ED'' and ``Discharge from the ED'' are accurate to the second level. Furthermore, the attribute ``temperature'' exhibits inconsistent data format. It is important to note that the majority of patient temperatures are recorded in degrees Fahrenheit, while some temperatures are documented in Celsius.

\paragraph{Remark} Based on the above validation, no errors were introduced during the log extraction process, and all the identified issues were inherited from the MIMIC-IV-ED dataset. Whether to handle these issues depends on specific analytical objectives. Refer to the next section for detailed examples. 


%% file: usage_notes.tex
\section*{Usage Notes}


This section presents analyses performed using \textbf{MIMICEL}, demonstrating the utility of the proposed dataset in addressing key questions related to the ED process. 
These analyses were inspired by frequently posed questions from ED experts on specific ED process activities, as summarised by Rojas et al.~\cite{rojas2017question}. These questions were obtained through a combination of interviews with ED specialists and literature reviews, ensuring that the questions reflect the needs for improving ED operations and management~\cite{rojas2017question}. Broadly, these questions are categorised into acuity-driven questions and ED  Length of stay (LoS)-oriented questions. Acuity-driven questions focus on the urgency of care, which is assessed during the triage stage, a critical step in ED operations where patients are assigned acuity levels to determine the priority of required interventions. 

Accordingly, the first analysis focuses on acuity, aiming to uncover characteristics and differences between ED processes for patients with varying acuity levels. 
The second analysis investigates ED LoS, a key performance indicator for evaluating ED efficiency~\cite{vanbrabant2019simulation}. This analysis seeks to provide a comprehensive understanding of the process characteristics associated with ED visits, distinguishing between normal and prolonged cases. 
As a key challenge in ED, overcrowding can lead to prolonged LoS~\cite{overcrowding}. Therefore, in the third analysis, we pay particular attention to the issue of overcrowding in ED. This analysis investigates the characteristics of ED processes under varying levels of crowdedness to better understand the impact of overcrowding on emergency care operations.

Based on insights from ED specialists, certain activities, such as ``Medicine reconciliation'', ``Medicine dispensation'', and ``Vital sign check'' may occur during transportation to the ED (e.g., via ambulance or helicopter) and are repeated upon the patient's arrival. In this work, we focus exclusively on ED activities that take place once the patient has entered the ED. Consequently, we filtered out events occurring at or prior to the event ``Enter the ED'', resulting in the removal of 80,581 events (out of 7,568,824 events), approximately 1.06\% of the total events in the \textbf{MIMICEL} dataset. All analyses presented were conducted using this filtered version of \textbf{MIMICEL}.


\subsection*{Acuity-based analysis}
In the MIMIC-IV-ED dataset, the assignment of acuity levels is guided by a five-level system of the Emergency Severity Index (ESI), which categorises patients based on the urgency of their condition. 
Level 1 represents the most urgent cases requiring immediate physician intervention, while level 2 patients are also at high risk, with their subsequent placement determined by nursing assessments. Patients at levels 3, 4, and 5 have gradually decreasing levels of urgency. 
The acuity-based analysis examines the influence of these acuity levels on ED process characteristics. During the validation of \textbf{MIMICEL}, it was identified that 1.64\% cases have missing values for the attribute \textit{acuity}. Consequently, these cases will not be covered in the following analyses.

\begin{figure}[h!!!]
    \centering
    \includegraphics[width=0.7\textwidth]{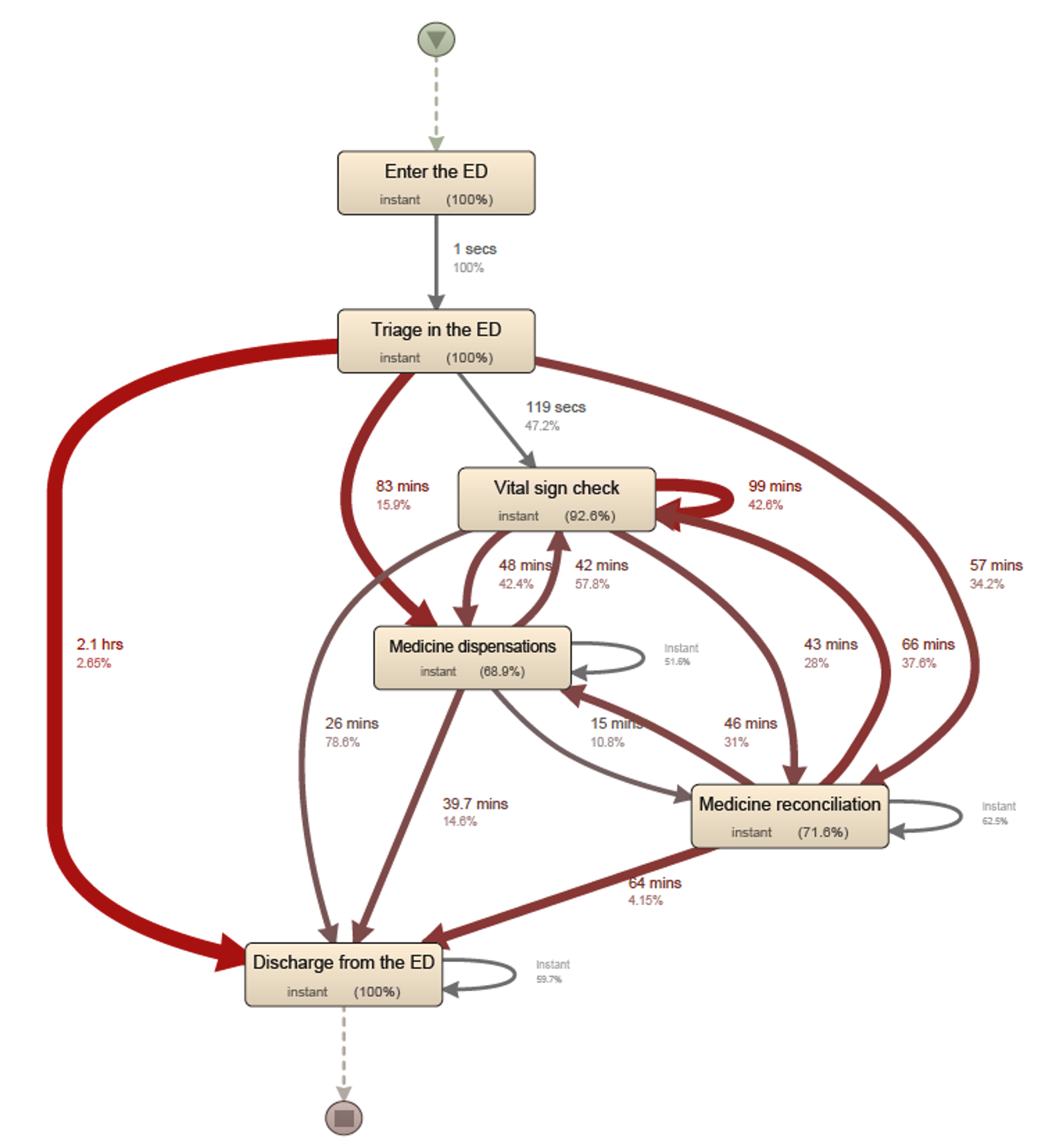}
    \caption{The process map of ED visits with the acuity level of 3}
    \label{fig:acuity3}
\end{figure}

With access to event logs, process maps can be generated to visualise an end-to-end process. For example, Figure~\ref{fig:acuity3} depicts a process map for ED visits with an acuity level of 3. This category accounts for over 50\% of cases in \textbf{MIMICEL}, making it a representative example. The process map, generated using the process mining tool called Disco\footnote{\url{https://fluxicon.com/disco/}}, provides a visualisation of the actual flow of the process for ED visits with this acuity level. Using colour and thickness coding, it highlights the relationships between activities, the most common paths (i.e., sequences of activities) taken by ED visits, and bottlenecks within the process. For instance, the path from triage to discharge is relatively uncommon, appearing in only 2.65\% of cases (with acuity level 3). The median time interval between these two activities is 2.1 hours (displayed by the thick red arrow), which indicates a bottleneck in this process. Additional details on interpreting the process map are provided in the Appendix. 

Table~\ref{tab:activity_frequency} summarises the case coverage of three optional activities across different acuity levels. Table~\ref{tab:path_frequency} presents the case coverage of paths observed across different acuity levels, and Table~\ref{tab:path_duration} provides the corresponding time interval between these activities. 

\begin{table}[h!]
\centering
\resizebox{.6\textwidth}{!}{%
\begin{tabular}{lccc}
\hline
                                                  & \multicolumn{3}{c}{Activity Observed in (\%) Cases}                \\ \cline{2-4} 
Acuity Level                                      & Medicine dispensation & Medicine reconciliation & Vital sign check \\ \hline
Acuity 1                                          & \textbf{81.5}         & 67.1                    & 93.4             \\
Acuity 2                                          & 72.3         & 78.7                    & 95.7             \\
Acuity 3                                          & \textbf{68.9}         & 71.6                    & 92.6             \\
Acuity 4                                          & 52.9         & 52.7                    & 82.6             \\
Acuity 5                                          & \textbf{27.6}         & 45.3                    & 68.3             \\ \hline
\end{tabular}%
}
\caption{The case coverage of three activities observed in cases with different acuity levels. A decreasing trend is observed in the frequency of ``Medicine dispensation'', ``Medicine reconciliation'' and ``Vital sign check'' as the urgent level decreases (from acuity 1 to acuity 5). ``Vital sign check'' remains the most consistently performed activity, which occurs in 68\% of cases even at acuity 5 cohort.
In contrast. ``Medicine dispensation'' and ``Medicine reconciliation'' are significantly less frequent at lower urgency levels, with a notable decline from acuity 3 onwards.}
\label{tab:activity_frequency}
\end{table}

\begin{table}[h!]
\centering
\resizebox{.7\textwidth}{!}{%
\begin{tabular}{lcccc}
\hline
                              & \multicolumn{4}{c}{\multirow{2}{*}{Path Frequency observed in (\%) Cases}}       \\
                              & \multicolumn{4}{c}{}                                                             \\ \cline{2-5} 
Acuity Level                  & Consec. Vital sign & Vital to Med Disp & Med Disp to Vital & Triage to Discharge \\ \hline
Acuity 1                      & \textbf{73.2}               & \textbf{65.2}              & \textbf{68.8}              & \textbf{1.23}                \\
Acuity 2                      & 65.2               & 56.5              & 63.2              & 1.35                \\
Acuity 3                      & 42.6               & 42.4              & 57.8              & 2.65                \\
Acuity 4                      & 15.7               & 18.8              & 37.7              & 6.89                \\
Acuity 5                      & \textbf{10.0}               & \textbf{8.9}              & \textbf{16.6}              & \textbf{18.00}               \\ \hline
\end{tabular}%
}
\caption{The case coverage of paths observed in cases with different acuity levels. As the urgency level decreases (from acuity 1 to acuity 5), the frequency of most paths declines, with the exception of the path from triage to discharge. Consecutive vital sign checks are the most frequently observed paths across all acuity levels, occurring in 73\% of cases for the acuity 1 cohort and dropping to 10\% for the acuity 5 cohorts. Similarly, paths between ``Vital sign check'' and ``Medicine dispensation'' follow this downward trend,  with higher frequencies at higher urgency levels (e.g., 65.2\% and 68.8\% for acuity 1) and significantly lower frequencies at reduced urgency levels (8.9\% and 16.6\% for acuity 5). In contrast, the path from triage to discharge, although the least common in urgent cases (e.g., 1.23\% for Acuity 1), becomes more frequent with decreasing urgency, reaching 18\% for acuity 5. }
\label{tab:path_frequency}
\end{table}

\begin{table}[h!]
\centering
\resizebox{.6\textwidth}{!}{%
\begin{tabular}{lcccc}
\hline
             & \multicolumn{4}{c}{Time Interval between Activities (min)}                       \\ \cline{2-5} 
Acuity Level & Consec. Vital sign & Vital to Med Disp & Med Disp to Vital & Triage to Discharge \\ \hline
Acuity 1     & \textbf{30}                 & \textbf{16}               & \textbf{14}                & \textbf{126}                 \\
Acuity 2     & 73                 & 38                & 28                & 126                 \\
Acuity 3     & 99                 & 48                & 42                & 126                 \\
Acuity 4     & 111                & 53                & 43                & 100                 \\
Acuity 5     & \textbf{120}                & \textbf{50}                & \textbf{36}                & \textbf{60}                  \\ \hline
\end{tabular}%
}
\caption{The Time Interval between activities observed in cases with different acuity levels. It is evident that the time intervals between activities generally increase as the urgency level decreases (from acuity 1 to acuity 5), except for the path from triage to discharge. The time interval for consecutive vital sign checks demonstrates a significant increase with decreasing urgency, commencing at 30 minutes for acuity 1 and escalating to 120 minutes for acuity 5. A similar upward trend is observed in the time intervals for the path between ``Vital sign check'' and ``Medication dispensation'', with a decrease in urgency. A notable exception is observed in the interval between triage and discharge, which exhibits a reverse trend, decreasing from 126 minutes for acuity 1 to 60 minutes for acuity 5.}
\label{tab:path_duration}
\end{table}

A comparative analysis of Table~\ref{tab:activity_frequency}, Table~\ref{tab:path_frequency} and Table~\ref{tab:path_duration} reveals the following insights:
\begin{itemize}
    \item \textbf{Higher-acuity patients require more frequent and intensive medical interventions} 

    The activity "Medicine dispensation" is significantly more common in higher-acuity cases, observed in 81.5\% of acuity 1 cases, gradually decreasing to 68.9\% for acuity 3 and further dropping to only 27.6\% in acuity 5 cases (Table~\ref{tab:activity_frequency}).
    This trend indicates that higher-acuity patients often require intensive medical interventions to stabilise their conditions, while lower-acuity patients frequently require minimal or no medication at all.
    
    \item \textbf{Lower-acuity patients require less frequent and less intensive vital sign monitoring} 
    
    ED visits with higher acuity levels require consecutive vital sign checks more frequently, reflecting the severity of patients' conditions. 
    For example, 73\% of cases in the acuity level 1 cohort involves consecutive monitoring of vital signs, compared to only 10\% in the acuity level 5 cohort (Table~\ref{tab:path_frequency}). Moreover, the interval between consecutive vital sign checks is significantly shorter for higher acuity levels, with a median duration of 30 minutes for acuity 1 and 120 minutes for acuity 5 (Table~\ref{tab:path_duration}). These findings suggest that as acuity decreases, patient conditions tend to be more stable, necessitating less frequent monitoring of their vital signs.
    
    \item \textbf{Higher-acuity patients require frequent and intensive cycles of medical interventions and vital sign monitoring} 
    
    ED visits with higher acuity undergo more frequent and rapid transitions between ``Medicine dispensation'' and ``Vital sign checks'', reflecting the intensive care required to manage their critical conditions. For example, in the acuity 1 cohort, 65.2\% of cases involve a transition from "vital sign check" to "medicine dispensation," and 68.8\% involve the reverse transition. In contrast, these proportions drop significantly for acuity 5, where only 8.9\% and 16.6\% of cases involve these transitions (Table~\ref{tab:path_frequency}). The median time intervals between these transitions are shorter for acuity 1 (16 and 14 minutes) compared to acuity 5 (50 and 36 minutes (Table~\ref{tab:path_duration}). These findings demonstrate that higher-acuity patients require frequent and tightly timed cycles of intervention, ensuring continuous monitoring and timely responses to their critical needs. Lower-acuity patients, on the other hand, experience fewer cycles with longer intervals, consistent with their more stable conditions.
    
    \item \textbf{Lower-acuity patients follow simpler and faster care pathways.} 
    
    Lower-acuity patients are more likely to be discharged directly after triage , as seen in 18\% of acuity 5 cases compared to only 1.23\% of acuity 1 cases (Table~\ref{tab:path_frequency}).
    In addition, the time interval between triage and discharge is shorter for the acuity 5 cohort (60 minutes) compared to acuity 1 (126 minutes) (Table~\ref{tab:path_duration}). This reflects the simplicity and efficiency of care for lower-acuity patients, in contrast to the prolonged pathways required for higher-acuity cases.
\end{itemize}

\subsection*{LoS-driven analysis}
\begin{figure}[h!]
    \centering
    \includegraphics[width=.68\textwidth]{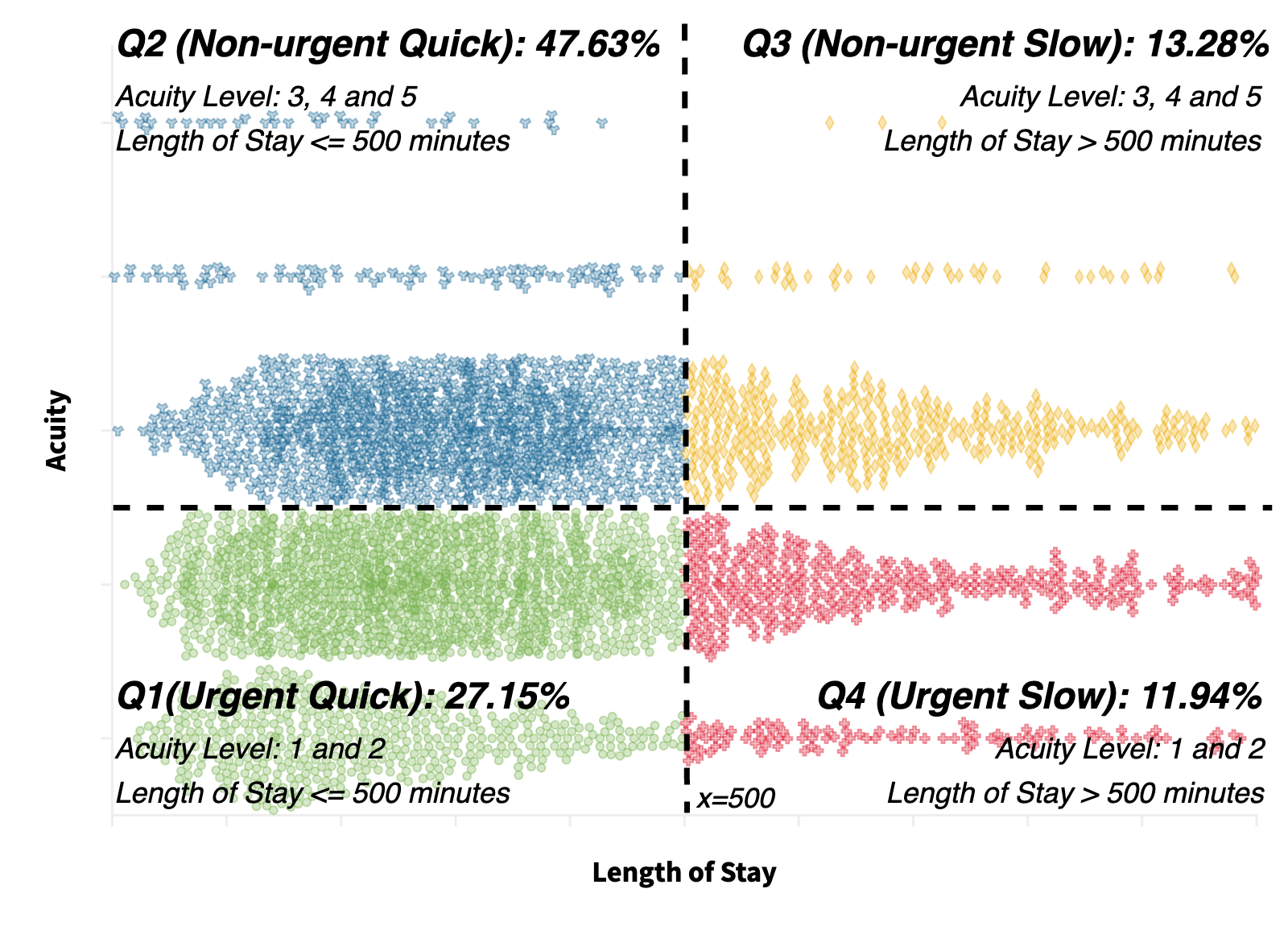}
    \vspace*{-\baselineskip}
    \caption{Distribution of ED visits in accordance with LoS and acuity levels}
    \label{fig:acuity-los}
\end{figure}

ED Length of Stay is an essential metric that measures the time between a patient's arrival at the ED and their physical departure from ED~\cite{vanbrabant2019simulation}. LoS represents the total duration of a patient’s stay in the ED and serves as a key indicator of ED efficiency. To explore factors contributing to prolonged LoS in ED, we incorporated the patient's acuity level. As previously discussed, acuity levels were assigned based on the five-level ESI system, where levels 1 and 2 represent high acuity and levels 3, 4, and 5 represent low acuity.
An analysis of the LoS distribution in the dataset reveals that 75\% of cases have a LoS of 500 minutes or less, which aligns with the internationally recommended acceptable ED LoS of \(\leq 8\) hours internationally~\cite{rose2012emergency}. Consequently, 500 minutes was established as the threshold for normal LoS. Cases with a LoS exceeding 500 minutes, comprising the remaining 25\% of cases, were classified as having a prolonged LoS. 
To facilitate the analysis, all cases were divided into four zones based on combinations of LoS (normal vs. prolonged) and acuity levels (high vs. low). Figure~\ref{fig:acuity-los} provides a visual representation of these zones, helping to identify distinct characteristics of processes with varying LoS and across different acuity levels. 

\begin{figure}[h!!]
    \vspace*{.25\baselineskip}
    \centering
    \includegraphics[width=\textwidth]{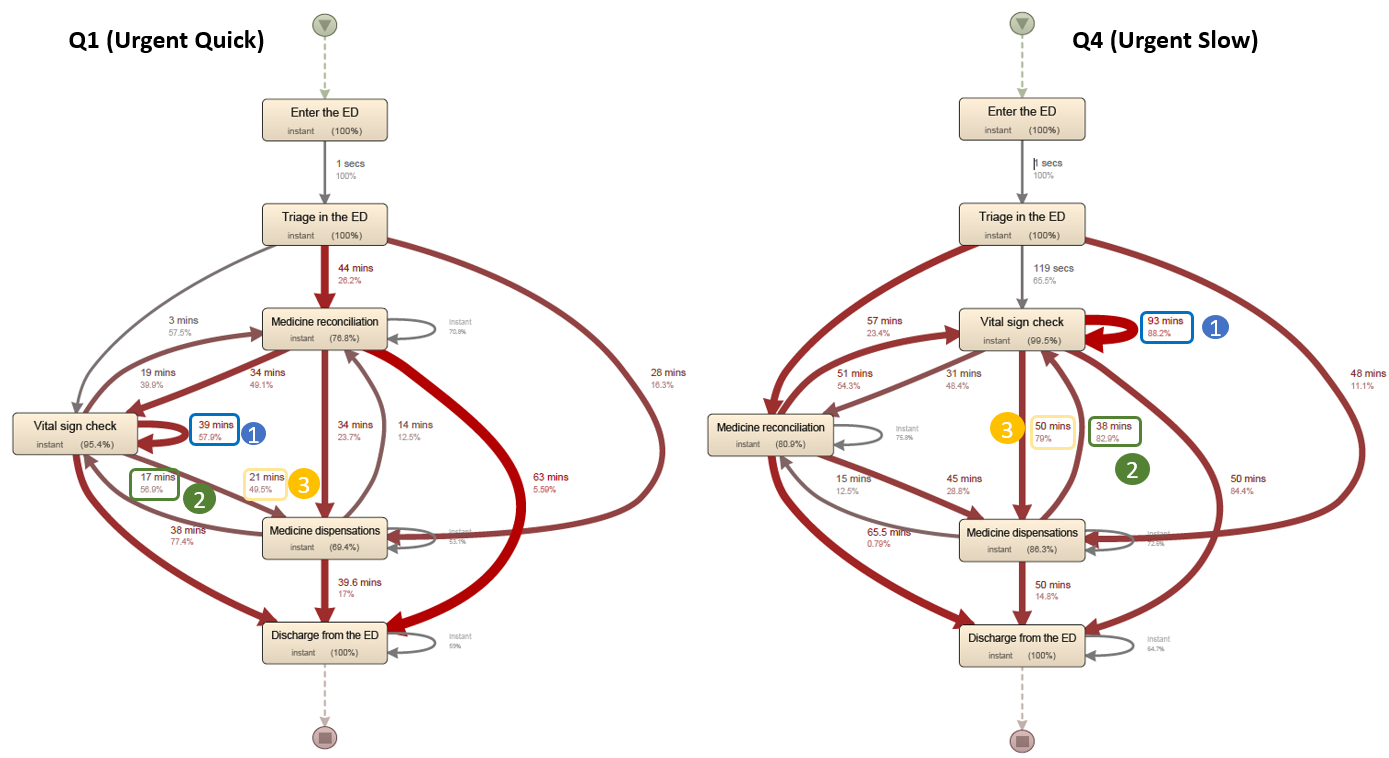}
    \caption{Comparison of the process between Q1 Urgent Quick zone and Q4 Urgent Slow zone}
    \label{fig:q1_vs_q4}
\end{figure}

 Based on this figure, we observed that approximately 27\% of cases fall into Q1, representing urgent cases with normal LoS, indicative of efficient management. In contrast, Q4, which accounts for 11.94\% of cases, represents urgent cases with prolonged LoS. The majority of cases, 47.63\%, fall into Q2, comprising non-urgent patients with shorter stays. A focused analysis of Q1 and Q4 enabled a detailed comparison of processes in normal versus prolonged LoS scenarios for urgent patients. 
The observations, illustrated in Figure~\ref{fig:q1_vs_q4} are summarised below:
\begin{enumerate}
    \item \textbf{Vital sign check self-loop } Consecutive vital sign checks occurred in 88\% of Q4 cases, compared to 58\% in Q1, indicating more intensive monitoring for patients with prolonged LoS.
    \item \textbf{Time duration taken by the path from ``Medicine dispensation'' to ``Vital sign check'' } The median duration of this path was twice as long as in Q1. Q4 also had a higher percentage of cases (83\%) with this path compared to 57\% in Q1. 
    \item \textbf{Time duration taken by the path from ``Vital sign check'' and ``Medicine dispensation'' } This path's duration was twice as long in Q4 compared to Q1. While 50\% of Q1 cases included this path, it was present in 80\% of Q4 cases.
\end{enumerate}

These observations indicate that Q4 patients, characterised by high acuity and prolonged LoS, experience significantly more intensive vital sign monitoring and slower transitions between medicine dispensation activities compared to Q1 patients. This suggests that the prolonged LoS in Q4 may result from the increased complexity and duration of ED processes required to stabilise these patients. 

Subsequently, we investigate the processes within the Q4 zone. Given that the majority of cases in \textbf{MIMICEL} are discharged either to ``Home'' (57\%) or ``Admitted to the hospital'' (37\%), we examine the processes with long LoS for these two dispositions. Consequently, we divide Q4 into two cohorts: the ``Home Cohort'' and the ``Admitted Cohort.'' Analysis revealed significant differences in process durations between these two groups, as shown in Table~\ref{tab:figure7_q4_durations}. 

\begin{table}[h!!!]
    \vspace*{.25\baselineskip}
    \centering
    \resizebox{\textwidth}{!}{%
\begin{tabular}{lrr}
\toprule
                                              Path &  Home Cohort (min) &  Admitted Cohort (min) \\
\midrule
Time duration of consecutive ``Vital sign check'' activities execution &                120 &                     64 \\
Time duration taken by the path from ``Vital sign check'' to ``Medicine dispensation'' &                 61 &                     39 \\
Time duration taken by the path from ``Medicine dispensation'' to ``Vital sign check'' &                 53 &                     24 \\
\bottomrule
\end{tabular}
}
\caption{Performance median duration (in minutes) of different paths for ED visits with different dispositions (Q4)}
\vspace*{-\baselineskip}
\label{tab:figure7_q4_durations}
\end{table}

\textit{Home Cohort vs Admitted Cohort}: For cases in the ``Home Cohort'', the median duration of consecutive vital sign checks was 120 minutes, nearly double the 64 minutes observed in the ``Admitted cohort''. Similarly, the duration of the path from ``Vital sign check'' to "Medicine dispensation" was longer for the ``Home Cohort'' (61 minutes) compared to the ``Admitted Cohort'' (39 minutes). A comparable trend was observed for the path from ``Medicine dispensation'' to ``Vital sign check'', with median durations of 53 minutes and 24 minutes for the ``Home'' and ``Admitted'' cohorts, respectively.
These observations suggest that patients discharged home experience more prolonged monitoring and slower transitions between activities, possibly reflecting extended observation times or additional checks to confirm their readiness for discharge. In contrast, patients in the ``Admitted cohort'' may follow a more streamlined process with quicker transitions, likely due to their immediate transfer to hospital wards for continued care. 


\subsection*{Crowdedness analysis}
ED overcrowding is a critical issue that arises when the demand for emergency healthcare services surpasses the capacity of an ED to deliver timely and suitable care to patients~\cite{sartini2022overcrowding}. In our dataset, we lack explicit information regarding the ED's capacity. Therefore, to address this limitation, we propose the establishment of crowdedness criteria derived from the statistical distribution of simultaneously treated patients within the ED. This criterion will indicate the level of overcrowding within the ED, allowing us to gain insights into the extent of the issue and its potential implications.

\paragraph{Technical definitions}

\begin{itemize}
\item {\it Simultaneously treated patients:} Let us consider an ED visit for a patient $p$, 
 whose enter and discharge times are denoted by $t_e$ and $t_d$, respectively. Any patient who entered the ED after $p$ was discharged, or any patient who was discharged before patient $p$ entered the ED, cannot be considered as simultaneously treated patients with patient $p$. Hence, a different patient $p'$ (with enter and discharge times $t'_e$ and $t'_d$) can only be considered as a simultaneously treated patient with $p$, only if the following logical statement holds: 

$p'$ is simultaneously treated with $p$ if $\neg (t'_e>t_d \lor t'_d<t_e)$   

\item{\it Crowdedness threshold:} For the purpose of crowdedness analysis, we determined the $75^{th}$ percentile of the distribution of the number of simultaneously treated patients in the ED as a threshold for determining ED crowdedness. If the number of simultaneous patients associated with a specific ED visit exceeds this threshold, we classify that patient visit as having taken place in a crowded ED. Applying this approach to \textbf{MIMICEL}, an ED is considered crowded when there are \textbf{12} or more simultaneously treated patients.
\end{itemize}


\paragraph{Crowded ED vs non-crowded (normal) ED}

\begin{figure}
    \centering
    \includegraphics[width=0.95\textwidth]{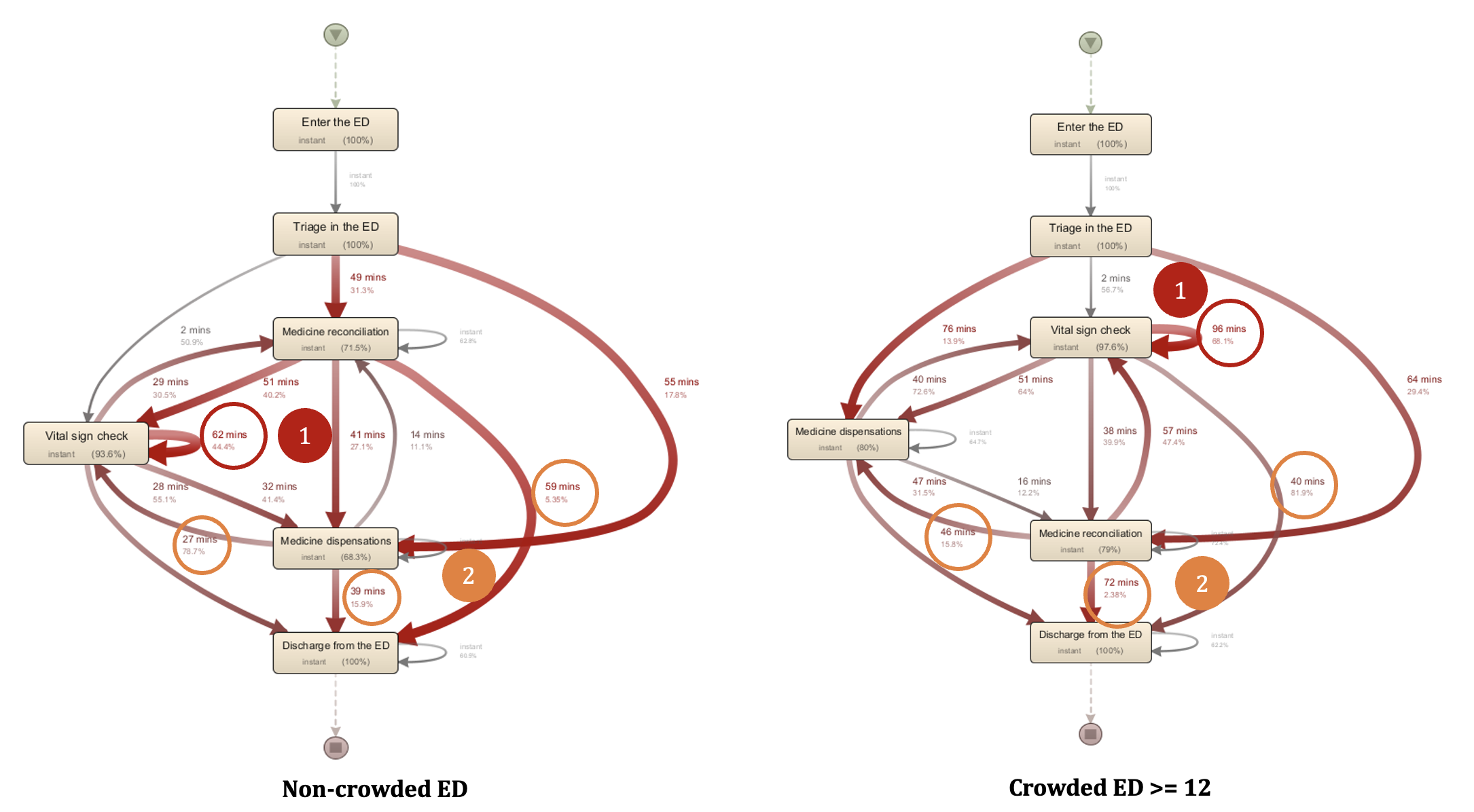}
    \caption{Comparison of the process between non-crowded and crowded ED}
    \label{fig:crowded-non-crowded-ed}
\end{figure}

Figure~\ref{fig:crowded-non-crowded-ed} depicts a comparison between crowded vs non-crowded ED. We analyse the process mainly in terms of time intervals between activities. The following observations can be obtained. 

\begin{enumerate}
    \item \textbf{Time interval between consecutive vital sign checks}: For the ``Home Cohort'', the time interval between consecutive vital sign checks is 2 hours compared to the 1-hour duration in the admitted cohort.
    \item \textbf{Time duration taken by the path from the second last activity 
    to ``Discharge from the ED''}: In a crowded ED, the time duration taken to the discharge from the immediate precedent activity is longer. The difference is primarily visible in the path from the ``Vital sign check'' to discharge and the path from ``Medicine Reconciliation'' to discharge.
\end{enumerate}

\textit{Home cohorts vs Admitted Cohort} An intriguing observation arose when the distribution of disposition cohorts was analysed between non-crowded and crowded ED. Within the entire ``Admitted Cohort'' consisting of 158,010 individuals, approximately one-third (48,156 patients) can be categorised as being treated in crowded ED conditions. On the other hand, within the ``Home Cohort'', which includes 241,626 patients, only one-fourth (60,384 patients) are associated with crowded ED conditions. These findings indicate that the proportion of admitted patients is significantly higher in crowded ED settings. Based on these findings, we conduct a more in-depth examination of the observations regarding time intervals between activities, specifically examining the intervals between consecutive vital sign checks and between the second last activity and discharge, with a focus on disparities between the ``Admitted Cohort'' and ``Home Cohort''.

Based on Table~\ref{tab:figure8_transition_durations}, we can observe that the longer time interval of consecutive vital sign checks is mainly contributed by the ``Home Cohort''. In terms of the time to discharge, the main difference between the two cohorts is visible in the path from the vital sign check to the discharge. In the ``Admitted Cohort'', this duration is considerably longer (58 minutes), compared to the ``Home Cohort'' (22 minutes). In the other discharge paths, the two cohorts display similar time intervals.


\begin{table}[h!]
\centering
\resizebox{\textwidth}{!}{%
\begin{tabular}{lrr}
\toprule
                                Path &  Crowded ED - Admitted (min) &  Crowded ED - Home (min) \\
\midrule
Time interval of consecutive ``Vital sign check'' activities &                           64 &                      120 \\
Time duration taken by the path from ``Vital sign check'' to ``Discharge from the ED''&                           58 &                       22 \\
Time duration taken by the path from ``Medicine reconciliation'' to ``Discharge from the ED''&                           68 &                       75 \\
Time duration taken by the path from ``Medicine dispensation'' to ``Discharge from the ED''&                           48 &                       38 \\
\bottomrule
\end{tabular}
}
\caption{Time interval differences between activities for Admitted and Home discharge cohorts in a crowded ED}
\label{tab:figure8_transition_durations}
\end{table}

%% file: code_availability.tex
\section*{Code Availability}

The code for data extraction, XES log conversion, technical validation, and analysis can be accessed online through our GitHub repository (\href{https://github.com/ZhipengHe/MIMIC-IV-event-log-extraction-for-ED}{https://github.com/ZhipengHe/MIMIC-IV-event-log-extraction-for-ED}). These scripts are publicly available to allow for reproducibility and code reuse. 

\begin{enumerate}
    \item Data Extraction (\textit{1\_extract\_eventlog}): This folder contains PostgreSQL scripts for extracting the event log from the MIMIC-IV-ED database and exporting them as CSV files. It is important to note that the \textbf{MIMICEL} event log generated in this study is derived from the MIMIC-IV-ED dataset~\cite{mimiced}. Consequently, valid access to the MIMIC-IV-ED dataset is required to use \textbf{MIMICEL}. The folder includes the following three SQL scripts:

    \begin{itemize}
        \item \textit{1\_preprocessing.sql}: preprocessing the MIMIC-IV-ED database and preparing for converting them to activities with timestamps.
        \item \textit{2\_to\_activity.sql}: converting the processed tables in the MIMIC-IV-ED database into activity tables.
        \item \textit{3\_to\_eventlog.sql}: combining all activity tables into a whole event log.
        \item \textit{4\_clean.sql}: cleaning invalid cases from event log where the attribute \textit{intime} is no earlier than \textit{outtime}. 
    \end{itemize}

    \item XES Log Conversion (\textit{2\_to\_xes}): This folder is running in a Python environment to convert the event log to the XES format. It includes a Python script and a Jupyter Notebook for log conversion:

    \begin{itemize}
        \item \textit{csv2xes.ipynb} \& \textit{csv2xes.py}: converting event log from CSV file to XES file.
    \end{itemize}

By the end of the data extraction and XES log conversion, the \textbf{MIMICEL} event log was obtained. This log serves as the foundation for subsequent analyses following a technical validation of its data quality.

    \item Technical Validation (\textit{3\_validation}): This folder includes details of the R package DaQAPO, which was utilised to assess the data quality of the event log. 
    \begin{itemize}
        \item \textit{data\_quality.Rmd}: detecting event log data quality issues, such as missing values, incomplete cases, violations of activity order, etc. (see details in the Technical Validation section)
        \item \textit{data\_quality.html} \& \textit{data\_quality\_revised.html}: storing the output report of technical validation. The revised version removes case lists in output report to improve the readability of HTML report.
    \end{itemize}


    \item Log Preparation (\textit{4\_analysis/log\_preparation}): This folder includes SQL scripts for filtering and generating meaningful insights from the event log:
    \begin{itemize}
        \item \textit{5\_insights.sql}: generating insights from the event log, such as the length of stay and static attributes.
        \item \textit{6\_filter.sql}: filtering the event log by removing events that happen before or have the same timestamp as ``Enter the ED'' as analyses conducted in this work focus exclusively on ED activities that take place onsite.
    \end{itemize}

    \item Log Analysis (\textit{4\_analysis/log\_analysis}) The extracted event log was used for further analysis with process mining tools and the Python environment. This analysis focused on:
    \begin{itemize}
        \item \textit{acuity\_cohorts.sql} \& \textit{throughput.sql}: extracting sublogs by acuity levels and discharge types for further analyses.
        \item \textit{acuity\_LoS.ipynb} \& \textit{crowdedness.ipynb}: demonstrating the usage of the filtered event log in the analysis (see details in the Usage Notes section).
    \end{itemize}
\end{enumerate}

%% file: Appendix.tex
\section*{Appendix}
\subsection*{Interpretation of process maps}

\begin{figure}[h!!!]
    \centering
    \includegraphics[width=\textwidth]{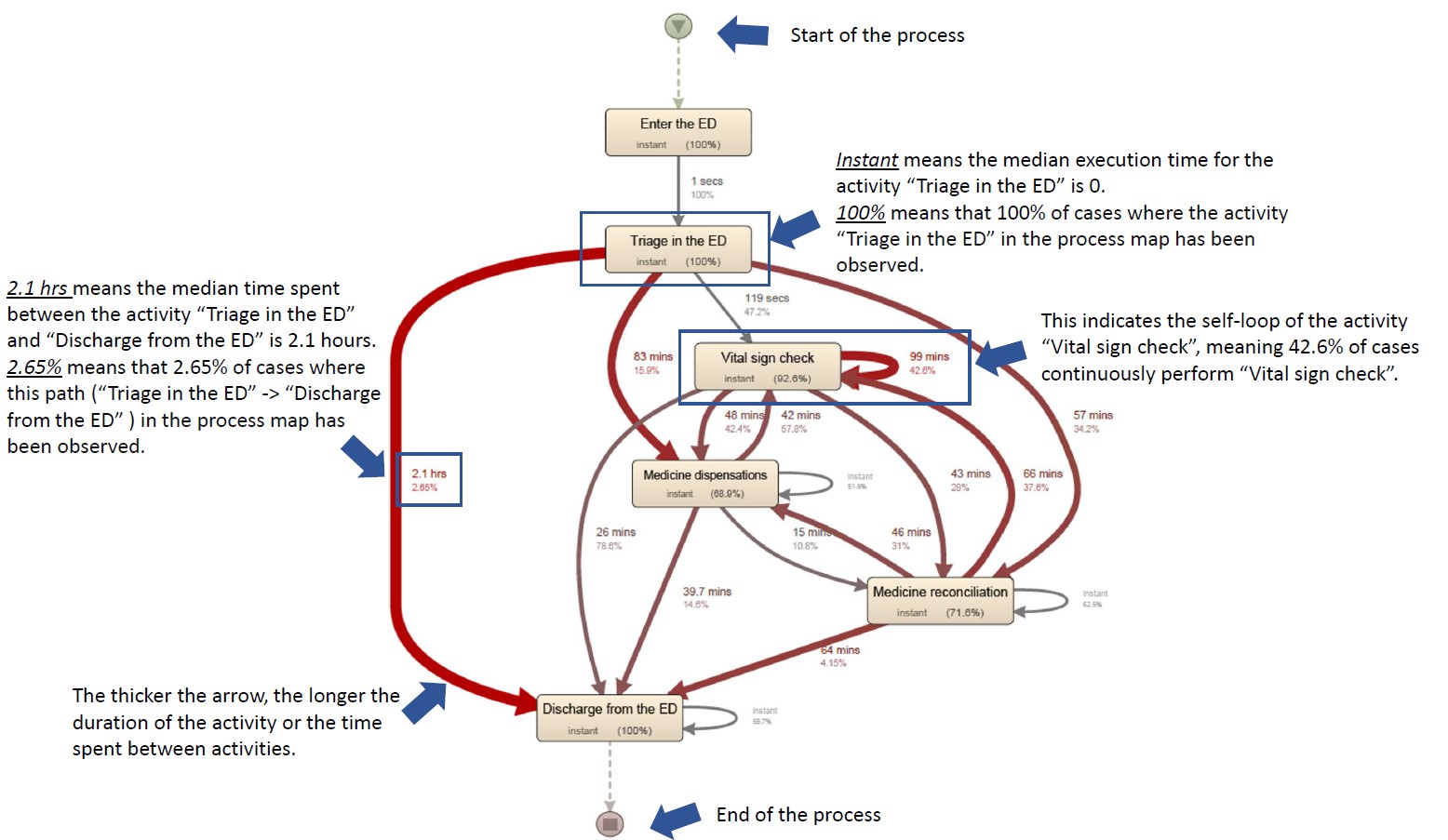}
    \caption{The annotated process map of ED visits with the acuity level of 3}
    \label{fig:acuity3_annotated}
\end{figure}

The process maps in this study were created using Disco, a process mining tool\footnote{\url{https://fluxicon.com/disco/}} designed for visualising and analysing processes discovered from event logs. To enhance the understanding of these process maps, we provide explanations of the notations used, offering essential background information. In the following, we outline the basic notations for a process map (as annotated in Figure~\ref{fig:acuity3_annotated}).
\begin{enumerate}
    \item \textbf{Process start: } The start of the process is represented by a triangle symbol located at the top of the process map. 
    \item \textbf{Process end: } The end of the process is denoted by a stop symbol.
    \item \textbf{Activities: } Activities are depicted as rectangle boxes. Each box include two elements:
    \begin{itemize}
        \item The median execution time of the activity. \item The percentage of cases that pass through the activity.
    \end{itemize}
    \item \textbf{Process flow: } The flow between two activities is represented by an arrow. 
    \begin{itemize} 
        \item Solid arrows indicate transitions between activities during the process. 
        \item Dashed arrows point to activities occurring at the very beginning or end of the process. \item Each arrow is annotated with two types of data: 
        \begin{itemize} 
            \item The median execution time of the path. \item The percentage of cases that follow the path. 
        \end{itemize} 
    \end{itemize}
\end{enumerate}

\newpage